\documentclass[10pt,doublecolumn]{IEEEtran}

\title{Ultra Low-Complexity Detection of Spectrum Holes in Compressed Wideband Spectrum Sensing}

\author{Zeinab Zeinalkhani and Amir H. Banihashemi \\
Department of Systems and Computer Engineering, Carleton University\\
Ottawa, Ontario, Canada\\
Email:\{zzeinab,ahashemi\}@sce.carleton.ca \vspace{-.0cm}}

\date{...}
\usepackage{color}
\usepackage{stmaryrd}
\usepackage{ifsym}
\usepackage{dsfont}
\usepackage{amssymb}
\usepackage{amssymb,amsmath}
\usepackage{graphicx}
\usepackage{caption}
\usepackage{subcaption}
\usepackage{cite}
\newcommand*{\Scale}[2][4]{\scalebox{#1}{$#2$}}%
\DeclareMathAlphabet{\mathpzc}{OT1}{pzc}{m}{it}

\DeclareMathSizes{10}{9.25}{5}{5}

\usepackage{geometry}
\begin{document}
\pagenumbering{gobble}
\maketitle
\begin{abstract}

Wideband spectrum sensing is a significant challenge in cognitive radios (CRs) due to requiring very high-speed analog-to-digital converters (ADCs), operating at or above the Nyquist rate.
Here, we propose a very low-complexity zero-block detection scheme that can detect a large fraction of spectrum holes from the sub-Nyquist samples, even when the undersampling ratio is very small.
The scheme is based on a block sparse sensing matrix, which is implemented through the design of a novel analog-to-information converter (AIC).
The proposed scheme identifies some measurements as being  zero and then verifies the sub-channels associated with them as being vacant. Analytical and simulation results are presented that demonstrate the effectiveness of the proposed method in reliable detection of spectrum holes with complexity much lower than existing schemes. This work also introduces a new paradigm in compressed sensing where one is interested in reliable detection of (some of the) zero blocks rather than the recovery of the whole block sparse signal.


\end{abstract}

\vspace{-.0cm}
\section{Introduction}

In a wireless communication environment, many of the primary users (PUs)
do not use their licensed frequency bands at all times. The surveys show that
the maximum frequency utilization of the allocated spectrum can be less than 10\% \cite{FCC}.
To increase the frequency utilization in such environments, secondary users (SUs)
equipped with cognitive radios (CRs)~\cite{MM99} can be deployed to use the available spectrum ({\em spectrum holes})
without interfering with the PUs. Spectrum sensing, as the task of finding the spectrum holes,
is thus one of the important functions performed by CRs.

As an integral part of spectrum sensing, an analog to digital converter (ADC) is often used to sample
the signal at or above the Nyquist rate. In wideband scenarios, this will require
very high-speed ADCs, which are challenging to implement.

One of the techniques of wideband spectrum sensing (WSS), which could relax the high-speed sampling requirement, is a channel-by-channel scanning approach. A tunable bandpass filter (BPF) scans one channel at a time. A narrowband spectrum sensing technique is then applied to sense if the scanned channel is vacant or not. This approach is, in general, slow and inflexible \cite{sweepTune}.
Another approach, called ``filter-bank''\cite{FB}, is based on using a parallel set of narrowband filters, each covering one channel. In this approach, however, an enormous number of RF components is required due to the parallel structure of the filter bank.

Since the spectrum is sparsely occupied, the information rate of the analog signal is much less than the Nyquist rate.
Therefore, an analog to information converter (AIC) can be used with sub-Nyquist sampling rates, based on the compressed sensing (CS) theory \cite{AIC,intro2cs}.

Compressed sensing is a signal processing technique that can recover an unknown
signal with high probability from a small set of linear projections,
called {\em measurements}, if the signal is sparse in some domain \cite{CS-baraniuk}.
Linear projections of the signal and its sparse representation are described by two matrices called \emph{measurement matrix} and \emph{sensing matrix}, respectively.

CS has been recently applied to the wideband spectrum sensing to alleviate the need for high sampling rates \cite{1,TS12,CSS4fixedFreq,ICC_SS}.
The spectrum is then reconstructed by a sparse signal recovery algorithm, see, e.g., \cite{1,TS12}.
To reduce the sampling rate, in \cite{CSS4fixedFreq,ICC_SS}, the block structure of the spectrum is also used in the reconstruction process.\footnote{block sparse signals are the sparse signals that have nonzero entries occurring in clusters.}
In all the above literature \cite{1,TS12,CSS4fixedFreq,ICC_SS}, the spectrum holes are detected from the reconstructed spectrum by applying an energy detection algorithm.
Although, in each CR, only finding one vacant sub-channel may be enough, in cooperative CR networks, finding more than one spectrum hole is of interest. Therefore, in this paper, our goal is to detect as many spectrum holes as possible with a given number of measurements. Neither the recovery of the signal, nor the detection of all spectrum holes is of direct interest. We thus introduce a new CS paradigm in which, one is interested in reliable detection of some of the zero blocks, even when the number of measurements is relatively low.

To the best of our knowledge, all wideband spectrum sensing methods in the literature that use CS for spectrum sensing (e.g., \cite{1,CSS4fixedFreq,ICC_SS}), are based on dense sensing matrices (matrices which have few, or no, zero entries). The recovery algorithms with such matrices are often based on linear or convex optimization with the complexity between $\mathcal{O}(N^2)$ and $\mathcal{O}(N^3)$, where $N$ is the dimension of the signal.
In mobile CRs with limited power and computational resources, it is impractical to use CS with such costly computations for the reconstruction of the spectrum.
In addition, the processing time required for signal reconstruction can impose a significant delay in the spectrum sensing.
These motivate us to look for an ultra low-complexity scheme for spectrum sensing with the goal of detecting a large proportion of spectrum holes.

A popular category of recovery algorithms for CS are those based on message-passing algorithms. In
particular, the approximate message-passing (AMP) algorithm of \cite{rev2-1}, which works with dense sensing matrices, has attracted much attention due to its remarkable
performance/complexity trade-off. In the context of block sparse signals,
it is demonstrated in \cite{DJM-2013} that AMP with James-Steins shrinkage
estimator (AMP-JS) can outperform the existing block sparse recovery algorithms.
Another subcategory of iterative message-passing recovery algorithms are verification-based ones \cite{sudocodes,pfister,yasser}, which are based on sparse sensing matrices and are thus of lower complexity compared to the message-passing algorithms on dense matrices (graphs). These algorithms however are very sensitive to noise. In addition, their application to the recovery of block sparse signals face a number of fundamental challenges.

In this paper, we apply an ultra low-complexity scheme to the wideband spectrum sensing, which is based on a block sparse sensing matrix.
To this end, we design a measurement matrix as part of the AIC, such that the resulting sensing matrix has a block sparse structure.
As will be seen later, the structure of the sensing matrix affects the performance of zero detection scheme. We analyze the proposed zero-block detection scheme for regular sensing matrices (those with the same number of non-zero blocks in each row and column) and then generalize the results to the case of irregular matrices.
We can optimize the structure of the sensing matrix to maximize the fraction of spectrum holes that can be detected for a given number of measurements.
We consider both noiseless and noisy scenarios in this work, and analyze the performance of the proposed algorithm in both cases. Our analytical and simulation results demonstrate the effectiveness of the proposed algorithm in reliable detection of spectrum holes (even when the number of measurements is relatively small) with complexity that is negligible compared to the existing block sparse recovery algorithms such as AMP-JS.

\vspace{-.1cm}
\section{Notations, Definitions and Ensembles of Sensing Matrices (Graphs)}
\label{sec:notations}

In this paper, we use capital bold letters to refer to matrices and lowercase bold letters for vectors.
We also use $\mathds{C}$ and $\mathds{R}$ to refer to the fields of complex and real numbers, respectively.
For the real and imaginary parts of a complex argument, we use $\Re(.)$ and $\Im(.)$, respectively.
In addition, $(.)^T$, $(.)^*$, $(.)^{-1}$ and $(.)^{\dag}$ denote the transpose, conjugate, matrix inversion and conjugate transpose, respectively.

The zero block detection scheme presented in this work can be described based on a bipartite graph representation of CS operation. Let $\mathcal{G}(\mathcal{V} \cup \mathcal{M},\mathcal{E}_{\mathcal{V}, \mathcal{M}})$ denote a bipartite graph where $\mathcal{V} \cup \mathcal{M}$ is the node set and $\mathcal{E}_{\mathcal{V}, \mathcal{M}}$ is the edge set, so that, every edge in $\mathcal{E}_{\mathcal{V}, \mathcal{M}}$ connects a node in $\mathcal{V}$ to a node in $\mathcal{M}$.
We refer to the sets $\mathcal{V}$ and $\mathcal{M}$ as \emph{variable nodes} (VNs) and \emph{measurement nodes} (MNs), respectively. 
In a bipartite graph, $\mathcal{E}_{S_1,S_2} \subseteq \mathcal{E}_{\mathcal{V}, \mathcal{M}}$ denotes the set of the edges that connect the set $S_1 \subseteq \mathcal{V}$ to the set $S_2 \subseteq \mathcal{M}$. In addition, in a bipartite graph, let $\mathcal{V}(m)$ and $\mathcal{M}(v)$ denote the set of VNs neighboring to the measurement node $m$ and the set of MNs neighboring to the variable node $v$, respectively.

Let $\mathcal{V}_z \subseteq \mathcal{V}$, $\mathcal{V}_{nz} \subseteq \mathcal{V}$, $\mathcal{M}_z \subseteq \mathcal{M}$ and $\mathcal{M}_{nz} \subseteq \mathcal{M}$ refer to the sets of zero VNs, non-zero VNs, zero MNs and non-zero MNs, respectively. Clearly, $\mathcal{V}_{nz}$ ($\mathcal{M}_{nz}$) is the complement of the set $\mathcal{V}_z$ ($\mathcal{M}_z$) with respect to $\mathcal{V}$ ($\mathcal{M}$).

In a bipartite graph, a node in $\mathcal{V}$ ($\mathcal{M}$) has degree $i$ if it is connected to $i$ nodes in $\mathcal{M}$ ($\mathcal{V}$).
We use $d(v)$ to denote the degree of $v$.
Let the polynomials $\lambda(x) = \sum_i \lambda_i x^i$ and $\rho(x) = \sum_i \rho_i x^i$, respectively, represent the degree distributions of VNs and MNs, where $\lambda_i$ and $\rho_i$ denote the fraction of degree-$i$ nodes in VNs and MNs, respectively. Clearly, $\lambda(1)=\rho(1)=1$ and $\bar{\lambda}|\mathcal{V}| = \bar{\rho}|\mathcal{M}| = |\mathcal{E}_{\mathcal{V}, \mathcal{M}}|$, where $\bar{\lambda} = \sum_i i \lambda_i$, $\bar{\rho} = \sum_i i \rho_i$ and $|.|$ denotes the cardinality of a set.

If the degree of all nodes in $\mathcal{M}$ and $\mathcal{V}$ are $d_M$ and $d_V$, respectively, the graph is called \emph{bi-regular} (\emph{regular}, in brief). Otherwise, it is called \emph{irregular}.
In a regular graph, $\rho(x) = x^{d_M}$ and $\lambda(x) = x^{d_V}$.

\vspace{-.0cm}
\section{System Model and Problem Statement}
\label{sec:SystemModel}

Suppose a cognitive radio receives and down-converts a wideband analog signal, $r(t)$, with the bandwidth of $W$.
We assume that $r(t)$ occupies $L$ consecutive, non-overlapping spectrum bands, each with bandwidth equal to $B = W/L$ Hz, referred to as {\em sub-channels}. The boundaries of the sub-channels are denoted by $f_0<f_1<\cdots <f_L$.

Suppose that the vector $\mathbf{r}\in {\mathds{R}}^{N}$ is the discrete representation of the analog signal $r(t)$, where $N$ corresponds to the Nyquist rate. Let $\bf x = F r$ be the spectrum of $\mathbf{r}$ with only $K<<N$ non-zero elements and $\bf F$ is the $N \times N$ unitary Fourier matrix.
Since $r(t)$ is a real signal, the spectrum $\bf x$ is conjugate symmetric. Therefore, we only focus on the positive frequency elements of $\bf x$, denoted by $^{+}\mathbf{x}$.

In the context of spectrum sensing, the spectrum is block sparse, i.e., at any given time, each sub-channel is occupied with probability $\alpha << 1$ independent of the other sub-channels. 
We refer to $\alpha$ as the \emph{sparsity ratio}.
Let $\mathbf{u}^{(i)}$ be the $i$th segment of ${}^{+}\mathbf{x}$, which corresponds to the elements in the range $[f_{i-1} , f_i)$, and $f_c^{(i)}=(f_{i}+f_{i-1})/2$ be the center frequency of the $i$th sub-channel.
Therefore, we can write ${}^{+}\mathbf{x}=[(\mathbf{u}^{(1)})^T,\cdots,(\mathbf{u}^{(L)})^T]^T$.
We also assume that the magnitude of non-zero elements of ${\bf x}$ follows a distribution, $\mathpzc{f}$. 

In the noiseless scenario, the measurement vector $\mathbf{y} \in \mathds{R}^{M}$ can be mathematically written as, $\bf y= \Phi r =   \Phi F^{-1} x =  \Theta x$, where $\bf \Theta$ is the sensing matrix and $\mathbf{\Phi} \in \mathds{R}^{M \times N}$ is the measurement matrix.
Since $\bf \Theta =\Phi F^{-1} $ and $\bf \Phi$ is a real matrix, each row of $\bf \Theta$ has conjugate symmetry.
Let $^+\mathbf{\Theta}$ denote the $M\times N/2 $ matrix including the last $N/2$ elements of each row of $\bf \Theta$.
In this paper, we assume that $\bf \Theta$ is a random block sparse matrix in row with $L$ blocks, in correspondence with the blocks of ${}^{+}\mathbf{x}$, and that only a few randomly selected blocks out of $L$ blocks are non-zero and the rests are zero.\footnote{The sparse ${}^{+}\mathbf{\Theta}$ keeps the complexity low and improves the performance of the recovery algorithm, as becomes evident later.} 
In Section \ref{sec:BLKAIC}, we will explain how a sensing matrix with this structure can be constructed.
Assume that the magnitudes of all elements of a non-zero block ${}^{+}\mathbf{\Theta}_m^{(l)}$ follow a distribution $\mathpzc{g}$, where ${}^{+}\mathbf{\Theta}_m^{(l)}$ be the $l$th block of the $m$th row of ${}^{+}\mathbf{\Theta}$. The $m$th element of $\bf y$ can be written as
\begin{eqnarray}
y_m = 2 \Re \left( \sum_{l=1}^L {}^{+}\mathbf{\Theta}_m^{(l)} . \mathbf{u}^{(l)} \right), \quad m = 1, \ldots, M,
\label{ym}
\end{eqnarray}

\noindent where ``.'' denotes the inner product.
Now, the problem is to detect the maximum number of spectrum holes from a given number of measurements $M$. Note that here, we are interested in very small values of $M$, i.e., $M<<L$. Such values of $M$ can be much smaller than typical values that are of interest in conventional CS frameworks where $M$ is small in comparison with $N$ (not $L$).

The relationships in (\ref{ym}) for $m =1, \ldots, M$, can be represented by a bipartite graph, referred to as a {\em sensing graph}, with $L$ VNs and $M$ MNs.
In this model, a VN is a sub-vector of $B=N/(2L)$ complex elements. 

\vspace{-.0cm}
\section{Design of an AIC with Block Sparse Sensing Matrix}
\label{sec:BLKAIC}

\begin{figure}[t]\vspace{-.2cm}
\centerline{\includegraphics[width=3.3in,height=2in]{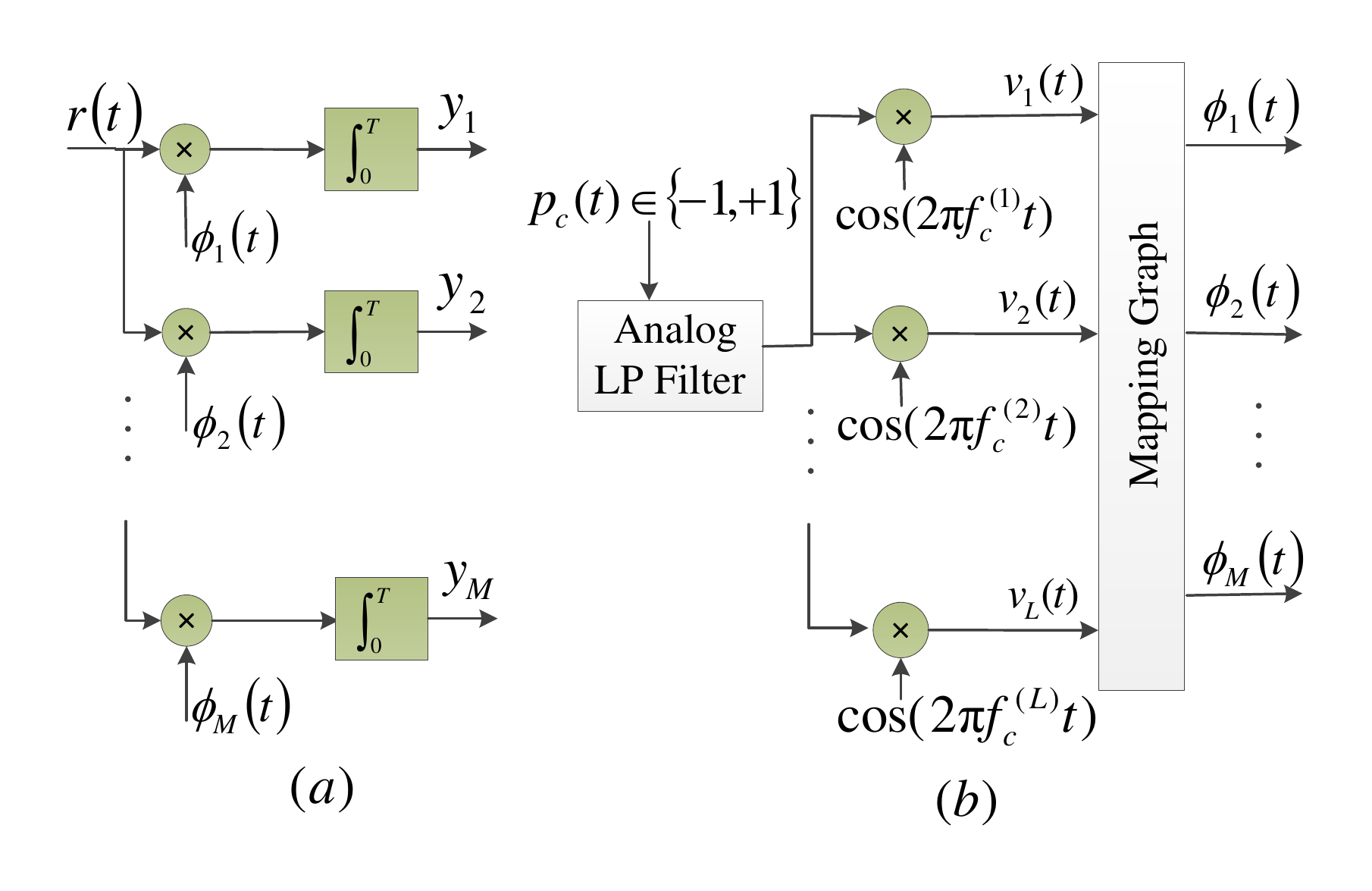}\vspace{-.2cm}}
\caption{{\small (a) The general structure of an AIC, (b) Sampling waveforms
of the AIC for a block sparse sensing.}\vspace{-.1cm}}
\label{fig:AIC_a_b_1}
\end{figure}

In Fig.~\ref{fig:AIC_a_b_1}(a), the general structure of an AIC with $M$ parallel branches of mixers and integrators (BMIs) is shown \cite{segmented-CS}.
In this section, the objective is to generate the sampling waveforms $\phi_i(t)$ of Fig.~\ref{fig:AIC_a_b_1}(a) to create a block sparse sensing matrix.

Suppose, $\mathbf{\Theta}_m$ and $\mathbf{\Phi}_m$ are the $m$th row-vectors of $\mathbf{\Theta}$ and $\mathbf{\Phi}$, respectively. We have $\mathbf{\Theta}_m^{\dag} =  (\mathbf{F}^{-1})^{\dag} \mathbf{\Phi}_m^{\dag}, \forall m$.
Since, $\mathbf{F}$ is a unitary matrix and $\mathbf{ \Phi}$ is a real matrix, we will have $\mathbf{ \Theta}_m^{\dag} =  \mathbf{F} \mathbf{\Phi}_m^{T}$, $\forall m$.
In other words, $\mathbf{\Theta}_m$ is the conjugate of the Fourier transform of $\mathbf{\Phi}_m$. Therefore, $\bf \Theta$ will have block sparse rows, when the rows of the measurement matrix are block sparse in the frequency domain.


Fig. \ref{fig:AIC_a_b_1}(b) demonstrates how to generate the sampling waveforms $\phi_i(t),\:i=1,\ldots,M$,
in Fig. \ref{fig:AIC_a_b_1}(a) to have a block sparse sensing matrix.
In Fig. \ref{fig:AIC_a_b_1}(b), the input signal, $p_c(t)$, is a noise generator with a rate equal to or higher than $W/L$.
Then, we filter the input signal by a low-pass (LP) filter with the cut-off frequency of $W/{2L}$ to result in a baseband signal with frequency range of $[-W/{2L},W/{2L}]$.
Finally, using the carrier frequencies $f_c^{(1)}, f_c^{(2)},\cdots,f_c^{(L)}$, the signal is modulated.
Now, we can generate $M$ sampling waveforms $\phi_m(t)$ using a linear combination of the resulted waveforms $v_l(t),\:l = 1, \ldots, L$, based on a properly designed mapping.
Note that a mapping with sparse linear combinations is needed to create a sparse sensing matrix. 

The mapping in Fig. \ref{fig:AIC_a_b_1}(b) can be constructed by the sensing graph described in Section \ref{sec:SystemModel}. Therefore, we can make a correspondence between the VNs and waveforms $v_l(t),\:l = 1, \ldots, L$, and between the MNs and sampling waveforms $\phi_m(t),\:m = 1, \ldots, M$, as follows
\[
\phi_m(t) = \sum_{l \in \mathcal{V}(m)} v_l(t) ,
\]
\noindent where $|\mathcal{V}(m)| = d(m)$.
Let $\tilde{\phi}_m(f)$ and $\tilde{v}_l(f)$ denote the Fourier transforms of $\phi_m(t)$ and $v_l(t)$, respectively. Therefore, $\tilde{\phi}_m(f) = \sum_{l \in \mathcal{V}(m)} \tilde{v}_l(f)$, i.e., $\mathbf{\Theta}_m$, which consists of the samples of $\tilde{\phi}^{*}_m(f)$, would be block sparse with $d(m)$ non-zero blocks in $^+\mathbf{\Theta}_m$.


In Fig. \ref{fig:AIC_a_b_1}(b), $L$ multipliers are required. 
We can generate the sampling waveforms using only $M$ multipliers, if we modulate the output signal of the LP filter by $\sum_{i \in \mathcal{V}(m)} \cos(2\pi f_c^{(i)}t),\:m = 1, \ldots, M$.

\vspace{-.1cm}
\section{Proposed Spectrum Hole Detection Scheme}
\label{sec:ZMD}

\subsection{Noiseless Scenario: Proposed Scheme and Its Analysis}

\noindent \emph{A.1. Proposed Scheme} \vspace{0.1cm}

The following lemma is the foundation of the proposed zero block detection in the absence of noise.

\textbf{Lemma 1:} If $y_m=0$ in (\ref{ym}), the sub-vectors of ${}^{+}\mathbf{x}$ corresponding to the non-zero blocks of ${}^{+}\mathbf{\Theta}_m$, are zero, with probability one.

\textbf{Proof}: 
The proof simply follows from the continuity of at least one the distributions $\mathpzc{f}$ or $\mathpzc{g}$, described in Section \ref{sec:SystemModel}. $\hfill \blacksquare$

Based on Lemma 1, the proposed scheme detects a zero block in the sensing graph, if that block is connected to a zero measurement node. We refer to this simple detection scheme for zero blocks (vacant sub-channels) as {\em zero measurement detection (ZMD)}.
In order to quantify the performance of the proposed method, we define two measures:

\begin{itemize}
\item[(i)] \emph{Probability of zero block detection} ($P_{ZD}$): {\small the ratio of the number of zero blocks detected correctly to the total number of zero blocks that exist in the spectrum.}
\item[(ii)] \emph{Probability of wrong zero block detection} ($P_{WZD}$): {\small the ratio of the number of occupied blocks detected falsely as vacant to the total number of the blocks detected as zero.}
\end{itemize}

Since $P_{WZD}$ indicates the level of interference to the PUs, it is an important measure from the standpoint of PUs.
Note that in the noiseless case, based on Lemma 1, $P_{WZD} = 0$ and the objective is to design a sensing graph in order to maximize $P_{ZD}$.

\vspace{0.3cm}
\noindent \emph{A.2. Analysis} \vspace{0.1cm}

In this section, we first consider bi-regular sensing graphs. In such graphs, the MN $m \in \mathcal{M}_z$, if and only if, all $d_M$ neighboring VNs belong to $\mathcal{V}_z$. Therefore, the probability that $m$ is zero is calculated by
\begin{eqnarray}
\textrm{P} \left(m \in {\cal{M}}_z \right) = \left(1-\alpha \right)^{d_M}.
\end{eqnarray}

\noindent Therefore, $E(|\mathcal{M}_z|) = M \left(1-\alpha \right)^{d_M}$.

On the other hand, the zero VN, $v$, is recovered by the ZMD method, if and only if, at least one of the neighboring MNs is in $\mathcal{M}_z$. By definition, $P_{ZD}$ is the ratio of the expected number of detected zero variables to the total number of zero VNs. Therefore, it is clear that $P_{ZD}$ is the probability that the zero VN, $v$, is recovered by ZMD, i.e., $P_{ZD}= \textrm{P} \left(|\mathcal{E}_{v,\mathcal{M}_z}| \geq 1 | v \in \mathcal{V}_z\right) = 1- \textrm{P} \left(|\mathcal{E}_{v,\mathcal{M}_z}|=0 | v \in \mathcal{V}_z\right)$.

For calculating this probability, let the edge $e$ connect a VN in $\mathcal{V}_z$ to a MN in $\mathcal{M}$. We first calculate the probability $p_0$ that the edge $e$ connects a VN in $\mathcal{V}_z$ to a MN in $\mathcal{M}_z$. Let $t_e$ and $h_e$ be the tail node and the head node of the edge $e$ in $\mathcal{M}$ and $\mathcal{V}$, respectively. Therefore, we can write
\begin{eqnarray}
p_0 &=& \textrm{P} \left(t_e \in \mathcal{M}_z | h_e \in \mathcal{V}_z\right)  \nonumber \\
&=& \frac{\textrm{P} \left(t_e \in \mathcal{M}_z , h_e \in \mathcal{V}_z\right)}{\textrm{P} \left(h_e \in \mathcal{V}_z\right)}
= \frac{\textrm{P} \left(t_e \in \mathcal{M}_z\right)}{\textrm{P} \left(h_e \in \mathcal{V}_z\right)}  \nonumber \\
&=&\frac{E\left(|\mathcal{E}_{\mathcal{V},\mathcal{M}_z}|\right)/|\mathcal{E}_{\mathcal{V},\mathcal{M}}|}{E\left(|\mathcal{E}_{\mathcal{V}_z,\mathcal{M}}|\right)/|\mathcal{E}_{\mathcal{V},\mathcal{M}}|}
= \frac{E\left(|\mathcal{E}_{\mathcal{V},\mathcal{M}_z}|\right)}{E\left(|\mathcal{E}_{\mathcal{V}_z,\mathcal{M}}|\right)}
= \frac{d_M E(|\mathcal{M}_z|)}{d_V |\mathcal{V}_z|} \nonumber \\
&=& \frac{d_M M \left(1-\alpha \right)^{d_M}}{d_V L(1-\alpha)} = \left(1-\alpha\right)^{d_M-1},
\end{eqnarray}
where the last equality comes from the fact that in bi-regular graphs, $d_V = M d_M / L$.
Thus, the probability that the zero variable node $v$ does not have any connection to $\mathcal{M}_z$, i.e., all $d_V$ neighboring MNs belong to $\mathcal{M}_{nz}$, is $(1-p_0)^{d_V}$. Therefore,
\begin{eqnarray}
P_{ZD} \hspace{-.25cm} &=& \hspace{-.25cm} 1 - \textrm{P}\left(|\mathcal{E}_{v,\mathcal{M}_z}|=0 | v \in \mathcal{V}_z \right)  \nonumber \\
\hspace{-.25cm} &=& \hspace{-.25cm} 1 - (1-p_0)^{d_V}  = 1 - \left( 1-\left(1-\alpha\right)^{d_M-1}\right) ^{d_V}.
\label{Pzd_reg}
\end{eqnarray}

We also can simply generalize the analysis of bi-regular sensing graphs to the case of irregular ones with the constraint that the VNs and MNs follow the degree distributions $\lambda(x)$ and $\rho(x)$, respectively.
After following the same steps as the ones for regular graphs, we can obtain $P_{ZD}$ as
\begin{eqnarray}
P_{ZD} = 1 - \sum_i \lambda_i (1-p'_0)^i,
\label{Pzd_irreg_wn}
\end{eqnarray}

\noindent where $p'_0 = {\sum_i i \rho_i \left(1-\alpha\right)^{i-1}}/{\sum_i i \rho_i}$.

\vspace{-.1cm}
\subsection{Noisy Scenario: Proposed Scheme and Its Analysis}
\label{sec:noisyMeasurements}

\noindent \emph{B.1. Proposed Scheme} \vspace{0.1cm}

In practice, the measurements are corrupted by noise, i.e.,
$\bf y= \Theta x  + n $, where $\mathbf{n}$ is an $M \times 1$ vector whose elements are independent and identically distributed (i.i.d.) zero-mean Gaussian random variables with variance $\sigma^2_n$, $\mathcal{N}(0,\sigma^2_n)$. Therefore, the $m$th measurement can be written as
\begin{eqnarray}
y_m = 2\Re\left(\sum_{i \in \mathcal{V}(m)} {}^{+}\mathbf{\Theta}_m^{(i)}.\mathbf{u}_i \right)+ n_m \:.
\label{noisyMNs}
\end{eqnarray}

Noise has a destructive effect on the application of ZMD method. In other words, if the noise-free measurement is equal to zero, its noisy version is not equal to zero with probability one. This will, in effect, disable ZMD method as described for noiseless scenarios. Here, we propose a test to detect zero measurements and we also show that in the high SNR regime, the test is translated to a thresholding approach, and
in this regime, we obtain the optimum threshold, analytically. Using the thresholding technique, the problem is reduced to one similar to the noiseless scenario. 

We consider an irregular sensing graph. For the analysis, we assume that the real and imaginary parts of non-zero elements of the sub-vectors $\mathbf{u}^{(i)}$ both come from a Gaussian distribution $\mathcal{N}(0,\sigma_s^2)$, in an i.i.d. fashion.
We consider $d(m)$ hypotheses $\mathcal{H}_i$ for $i=0,\ldots,d(m)$, where $\mathcal{H}_i$ represents the scenario that the measurement $m$ is connected to $i$ non-zero blocks and $d(m)-i$ zero-blocks. Therefore, the a priori probability of $\mathcal{H}_i$ is
\begin{eqnarray}
\textrm{P}\left(\mathcal{H}_i\right) =  {d(m) \choose i} \alpha^i (1-\alpha)^{d(m) - i} , \quad i=0,\ldots,d(m).
\label{pdf_Hi}
\end{eqnarray}



Furthermore, suppose that only one of the neighboring blocks of the $m$th MN (e.g., the $j$th block) out of $d(m)$ blocks is non-zero and $d(m)-1$ neighboring blocks are zero (i.e., Hypothesis $\mathcal{H}_1$ is true). In this case, we have
\begin{eqnarray}
y_m \hspace{-.3cm} &=& \hspace{-.3cm} 2\Re\left({}^{+}\mathbf{\Theta}_m^{(j)} . \mathbf{u}^{(j)} \right)+ n_m \nonumber \\
\hspace{-.3cm} &=& \hspace{-.3cm} 2 \sum_{i=1}^{D} { \Re({}^{+}\mathbf{\Theta}_{m,i}^{(j)}) . \Re(\mathbf{u}_i^{(j)}) - \Im({}^{+}\mathbf{\Theta}_{m,i}^{(j)}) . \Im(\mathbf{u}_i^{(j)}) } + n_m,
\label{noisyMN_exVersion}
\end{eqnarray}

\noindent where ${}^{+}\mathbf{\Theta}_{m,i}^{(j)}$ and $\mathbf{u}_i^{(j)}$ are the $i$th element of the sub-vector ${}^{+}\mathbf{\Theta}_{m}^{(j)}$ and the $i$th element of the sub-vector $\mathbf{u}^{(j)}$, respectively.  The probability density function of $y_m$ under this assumption ($\mathcal{H}_1$) is
\begin{eqnarray}
\textrm{P}(y_m|\mathcal{H}_1) = \mathcal{N}\left(0,4\|{}^{+}\mathbf{\Theta}_{m}^{(j)}\|_2^2 \sigma_s^2 +\sigma_n^2 \right).
\label{Pdf_yGivH1}
\end{eqnarray}

Based on the structure of the AIC presented in Section \ref{sec:BLKAIC}, $\|{}^{+}\mathbf{\Theta}_{m}^{(j)}\|_2^2 = cte,\ \forall j$. Without loss of generality, we assume that $\|{}^{+}\mathbf{\Theta}_{m}^{(j)}\|_2^2 = 1$. Therefore, $\textrm{P}(y_m|\mathcal{H}_1) = \mathcal{N}\left(0,4 \sigma_s^2 +\sigma_n^2 \right)$. Similarly, the probability density function of $y_m$ under the hypothesis $\mathcal{H}_i$ is given by
\begin{eqnarray}
\textrm{P}(y_m|\mathcal{H}_i) = \mathcal{N}\left(0,4 i \sigma_s^2 +\sigma_n^2 \right) , \quad i=0,\ldots,d(m).
\label{Pdf_yGivHi}
\end{eqnarray}


In ZMD, we need to determine whether the noise-free version of $y_m$, i.e., $y_m^{nf}$, is zero or not. If $y_m^{nf}$ is zero, $\mathcal{H}_0$ is true with probability one. Otherwise, $\mathcal{H}_i$ for $i=1,\ldots,d(m)$ is true ($\bar{\mathcal{H}}_0$ is true), i.e., $\mathcal{H}_0$ is false.
To make a decision on whether $y^{nf}_m$ is zero or not, we use the likelihood ratio test (LRT). The LR is defined by
\begin{eqnarray}
\Lambda(y_m) = \frac{\textrm{P}(y_m|\mathcal{\bar{H}}_0)}{\textrm{P}(y_m|\mathcal{H}_0)},
\label{LRT_formula}
\end{eqnarray}

\noindent and the test is performed as follows:
\begin{eqnarray}
\textrm{ if } \Lambda(y_m) < c &:&  \textrm{ accept } \mathcal{H}_0; \nonumber \\
\textrm{ if } \Lambda(y_m) \geq c &:&  \textrm{ reject } \mathcal{H}_0 \textrm{ (accept $\mathcal{\bar{H}}_0$)},
\label{test}
\end{eqnarray}

\noindent where $c$ is a properly selected threshold.

We can calculate $\textrm{P}(y_m|\mathcal{\bar{H}}_0)$ in (\ref{LRT_formula}) by
%
\begin{eqnarray}
\textrm{P}(y_m|\mathcal{\bar{H}}_0) = \frac{1}{\textrm{P}(\mathcal{\bar{H}}_0)} \sum_{i=1}^{d(m)} \textrm{P}(y_m|\mathcal{H}_i) \textrm{P}(\mathcal{H}_i).
\label{PYgH1}
\end{eqnarray}

Therefore, using (\ref{pdf_Hi}), (\ref{Pdf_yGivHi}), (\ref{LRT_formula}) and (\ref{PYgH1}), we have
\begin{eqnarray}
 \Lambda\hspace{-.32cm} &(& \hspace{-.32cm} y_m) = \frac{\sum_{i=1}^{d(m)} {d(m) \choose i} \alpha^i (1-\alpha)^{d(m) - i} \mathcal{N}\left(0,4i \sigma_s^2+\sigma_n^2 \right)}{\sum_{i=1}^{d(m)} {d(m) \choose i} \alpha^i (1-\alpha)^{d(m) - i} \mathcal{N}\left(0,\sigma_n^2\right)} \nonumber \\
 \hspace{-.25cm}&=& \hspace{-.25cm} \frac{\sum_{i=1}^{d(m)} {d(m) \choose i} \alpha^i (1-\alpha)^{d(m) - i} \sqrt{\frac{\sigma_n^2}{4i \sigma_s^2+\sigma_n^2}} \exp\{\frac{2i \sigma_s^2 y_m^2}{(4i \sigma_s^2+\sigma_n^2)\sigma_n^2}\}}{1-(1-\alpha)^{d(m)}}. \nonumber \\
 & &
 \label{LRT}
\end{eqnarray}

In the high SNR regime, where $\sigma_s^2 >> \sigma_n^2$, we can neglect $\sigma_n^2$ against $\sigma_s^2$ in (\ref{LRT}). We can thus simplify the decision for $\mathcal{H}_0$ (i.e., $\Lambda(y_m) < c$) to
\begin{eqnarray}
|y_m| < c',
\label{ZM_thr}
\end{eqnarray}

\noindent where $c'$ is determined by the target $P_{WZD}$ as described in the next section.
In other words, if $|y_m| < c'$, we decide $\hat{y}_m^{nf} = 0$ and if $|y_m| \geq c'$, we decide $\hat{y}_m^{nf} \neq 0$. 

\vspace{0.2cm}
\noindent \emph{B.2. Analysis} \vspace{0.1cm}

Based on the likelihood ratio test, the measurement values are partitioned into two regions:
\begin{eqnarray}
\mathcal{R}_0 = \{y_m : |y_m|<c'\}, \textrm{ and } \mathcal{R}_1 = \{y_m : |y_m| \geq c'\}.
\end{eqnarray}

Based on the correct or the erroneous detection of zero/non-zero measurements, we can partition the set of measurements $\mathcal{M}$ into four different sets: $\mathcal{M}_z^c$, $\mathcal{M}_z^w$, $\mathcal{M}_{nz}^c$ and $\mathcal{M}_{nz}^w$.
The set $\mathcal{M}_z^w$ represents the set of non-zero measurements which are detected erroneously as zero, $\mathcal{M}_z^w = \{y_m: y_m \in \mathcal{R}_0 , \mathcal{\bar{H}}_0\}$. Similarly, $\mathcal{M}_z^c = \{y_m: y_m \in \mathcal{R}_0 , \mathcal{H}_0\}$, $\mathcal{M}_{nz}^w = \{y_m: y_m \notin \mathcal{R}_0 , \mathcal{H}_0\}$ and $\mathcal{M}_{nz}^c = \{y_m: y_m \notin \mathcal{R}_0 , \mathcal{\bar{H}}_0\}$.
In the following, we first consider the case where the sensing graph is regular.

The probability of false alarm (for zero measurements) $P_{FA}$ is defined as the probability that a measurement $y_m$ is detected as zero while at least one of its neighboring VNs belongs to $\mathcal{V}_{nz}$, i.e., $P_{FA} = \int_{y \in \mathcal{R}_0} \textrm{P}(y| \mathcal{\bar{H}}_0) dy$. For the signal model considered here, by using (\ref{PYgH1}), we have

\begin{eqnarray}
P_{FA} \hspace{-.25cm} &=& \hspace{-.25cm}
\frac{\sum_{i=1}^{d_M} {d_M \choose i} \alpha^i (1-\alpha)^{d_M - i} \int_{y \in \mathcal{R}_0} \textrm{P}(y| \mathcal{H}_i) dy}{1-(1-\alpha)^{d(m)}} \nonumber \\
\hspace{-.25cm} &=& \hspace{-.25cm} \frac{\sum_{i=1}^{d_M} {d_M \choose i} \alpha^i (1-\alpha)^{d_M - i} \textrm{erf}(\frac{c'}{\sqrt{2(4i\sigma_s^2+\sigma_n^2)}})}{1-(1-\alpha)^{d(m)}},
\end{eqnarray}

\noindent where $\textrm{erf}(x) := \left(2/\sqrt{\pi}\right) \int_0^x e^{-t^2} dt$.

Probability of detection (for zero measurements), $P_D$, is defined as the probability that a zero measurement is detected correctly as zero, i.e., $P_{D} = \int_{y \in \mathcal{R}_0} \textrm{P}(y| \mathcal{H}_0)dy$. For our signal model, $P_{D}$ is given by
\begin{eqnarray}
P_{D} = \textrm{erf}(\frac{c'}{\sqrt{2\sigma_n^2}}).
\end{eqnarray}

Clearly, $E(|\mathcal{M}_z^c|) = M \textrm{P}(\mathcal{H}_0) P_{D} = M (1-\alpha)^{d_M} P_{D}$ and $E(|\mathcal{M}_z^w|) = M \textrm{P}(\mathcal{\bar{H}}_0)P_{FA} = M (1-\textrm{P}(\mathcal{H}_0)) P_{FA} = M (1-(1-\alpha)^{d_M}) P_{FA} $.

Now, we obtain $P_{WZD}$ and $P_{ZD}$ of the VNs based on the $P_{D}$ and $P_{FA}$ of the MNs. Based on the definition, we have 
\begin{eqnarray}
P_{WZD}
\hspace{-.25cm} &=& \hspace{-.25cm} \textrm{P}\left(v \in \mathcal{V}_{nz} | |\mathcal{E}_{v,\mathcal{M}_z^c\cup \mathcal{M}_z^w}|\geq 1\right) \nonumber \\
\hspace{-.25cm} &=& \hspace{-.25cm} \frac{\textrm{P}\left(|\mathcal{E}_{v,\mathcal{M}_z^c\cup \mathcal{M}_z^w}| \geq 1 | v \in \mathcal{V}_{nz}\right) \textrm{P}\left(v \in \mathcal{V}_{nz}\right)}{\textrm{P}\left(|\mathcal{E}_{v,\mathcal{M}_z^c\cup \mathcal{M}_z^w}| \geq 1\right)}.
\label{Pwzd_1}
\end{eqnarray}

Since, with probability one, $\mathcal{E}_{v,\mathcal{M}_z^c} = \emptyset$ for $v \in \mathcal{V}_{nz}$, we can write
\begin{eqnarray}
\textrm{P}\hspace{-.45cm} & & \hspace{-.45cm} \left(|\mathcal{E}_{v,\mathcal{M}_z^c\cup \mathcal{M}_z^w}| \geq 1 | v \in \mathcal{V}_{nz}\right)
= \textrm{P}\left(|\mathcal{E}_{v,\mathcal{M}_z^w}| \geq 1 | v \in \mathcal{V}_{nz}\right) \nonumber \\
\hspace{-.25cm} &=& \hspace{-.25cm} 1 - \textrm{P}\left(|\mathcal{E}_{v,\mathcal{M}_z^w}| =0 | v \in \mathcal{V}_{nz}\right)  \nonumber \\
\hspace{-.25cm} &=& \hspace{-.25cm} 1-\left(1-\textrm{P}\left(t_e \in \mathcal{M}^w_z | h_e \in \mathcal{V}_{nz} \right)\right)^{d_V},
\label{Pwzd_num1}
\end{eqnarray}
\noindent where $\textrm{P}\left(t_e \in \mathcal{M}^w_z | h_e \in \mathcal{V}_{nz} \right) =P_{FA}$.

For the denominator of (\ref{Pwzd_1}), we have
\begin{eqnarray}
\textrm{P}\left(|\mathcal{E}_{v,\mathcal{M}_z^c\cup \mathcal{M}_z^w}| \geq 1\right) \hspace{-.25cm} &=& \hspace{-.25cm} 1 - \textrm{P}\left(|\mathcal{E}_{v,\mathcal{M}_z^c\cup \mathcal{M}_z^w}| = 0 \right) \nonumber \\
\hspace{-.25cm} &=& \hspace{-.25cm}  1- \left(1 - \textrm{P}\left(t_e \in \mathcal{M}_z^c \cup \mathcal{M}_z^w \right) \right)^{d_V} ,
\label{Pwzd_denum}
\end{eqnarray}
\noindent where
\begin{eqnarray}
\textrm{P}\left(t_e \in \mathcal{M}_z^c \cup \mathcal{M}_z^w \right) \hspace{-.25cm} &=& \hspace{-.25cm} \frac{E(|\mathcal{M}_z^c|)+E(|\mathcal{M}_z^w|)}{M} \nonumber \\
\hspace{-.25cm} & & \hspace{-2cm} = (1-\alpha)^{d_M} P_D + (1-(1-\alpha)^{d_M}) P_{FA}.
\end{eqnarray}

Since, $\textrm{P} \left(v \in \mathcal{V}_{nz}\right) = \alpha$, from (\ref{Pwzd_num1}) and (\ref{Pwzd_denum}), $P_{WZD}$ is calculated by
\begin{eqnarray}
P_{WZD} \hspace{-.25cm} &=& \hspace{-.25cm} \alpha \frac{1-\left(1- P_{FA}\right)^{d_V} }{1-\big(1-(1-\alpha)^{d_M} P_{D}-(1-(1-\alpha)^{d_M}) P_{FA}\big)^{d_V}} \:.\nonumber \\
&&
\label{Pwzd}
\end{eqnarray}
\vspace{-.5cm}

The probability of zero detection, $P_{ZD}$, can be calculated by
\begin{eqnarray}
P_{ZD} \hspace{-.25cm} &=& \hspace{-.25cm} \textrm{P}\left(|\mathcal{E}_{v,\mathcal{M}_z^c\cup \mathcal{M}_z^w}| \geq 1 | v \in \mathcal{V}_z \right) \nonumber \\
\hspace{-.25cm} &=& \hspace{-.25cm} 1 - \textrm{P}\left(|\mathcal{E}_{v,\mathcal{M}_z^c\cup \mathcal{M}_z^w}| =0 | v \in \mathcal{V}_z \right) \nonumber \\
\hspace{-.25cm} &=& \hspace{-.25cm} 1- \left(1 - \textrm{P}\left(t_e \in \mathcal{M}_z^c\cup \mathcal{M}_z^w | h_e \in \mathcal{V}_z \right) \right)^{d_V} ,
\end{eqnarray}

\noindent where
\begin{eqnarray}
\textrm{P}\hspace{-.45cm}&&\hspace{-.45cm}\left(t_e \in \mathcal{M}_z^c\cup \mathcal{M}_z^w | h_e \in \mathcal{V}_z \right) = \frac{\textrm{P}\left(t_e \in \mathcal{M}_z^c\cup \mathcal{M}_z^w , h_e \in \mathcal{V}_z \right)}{\textrm{P}\left( h_e \in \mathcal{V}_z \right)} \nonumber \\
\hspace{-.35cm} &=& \hspace{-.25cm} \frac{E(|\mathcal{E}_{\mathcal{V}_z,\mathcal{M}_z^c}|)+E(|\mathcal{E}_{\mathcal{V}_z,\mathcal{M}_z^w}|)}{E(|\mathcal{E}_{\mathcal{V},\mathcal{M}}|)}.\frac{1}{(1-\alpha)} \nonumber \\
\hspace{-.35cm} &=& \hspace{-.25cm} \Scale[1.15]{\frac{E(|\mathcal{M}^c_z|)d_M + E(|\mathcal{M}^w_z|) \sum_{i=1}^{d_M} (d_M-i) {d_M \choose i} \alpha^{i} (1-\alpha)^{d_M-i}/\textrm{P}(\mathcal{\bar{H}}_0)}{M d_M (1-\alpha)}} \nonumber \\
\hspace{-.35cm} &=& \hspace{-.25cm} (1-\alpha)^{d_M-1} P_{D}+ (1-(1-\alpha)^{d_M-1})P_{FA}.
\end{eqnarray}

\noindent Therefore,
\begin{eqnarray}
\Scale[.93]{P_{ZD}= 1- \bigg(1 - (1-\alpha)^{d_M-1} \hspace{-.1cm} P_{D} - \left(1-(1-\alpha)^{d_M-1}\right)P_{FA}\bigg)^{d_V}.}
\label{Pzd}
\end{eqnarray}

When the measurements are noiseless, $P_{D} = 1$ and $P_{FA} = 0$. With these values for $P_{D}$ and $P_{FA}$, $P_{WZD}$ in (\ref{Pwzd}) will be zero and $P_{ZD}$ in (\ref{Pzd}) is reduced to (\ref{Pzd_reg}). With a large value for the threshold $c'$, all the measurements are detected as zero, which results in $P_D=1$ and $P_{FA}=1$. We see from Equations (\ref{Pwzd}) and (\ref{Pzd}) that, in this case, $P_{WZD}=\alpha$ and $P_{ZD}=1$, which are expected.

For given values of $d_M$, $d_V$, $\sigma_s$ and $\sigma_n$, the threshold $c'$ can be chosen to provide a target $P_{WZD}$. We can then obtain $P_{ZD}$ using this threshold level.
It is important to note that while $P_D$ and $P_{FA}$ depend on the signal model, $P_{WZD}$ and $P_{ZD}$ do not. Therefore, Equations (\ref{Pwzd}) and (\ref{Pzd}) can be applied to other signal models.


The above results can be generalized to irregular sensing graphs (matrices). The derivation of $P_{ZD}$ and $P_{WZD}$ is similar to (\ref{Pwzd_1})-(\ref{Pzd}). 
For irregular sensing matrices, we have
\begin{eqnarray*}
P_{WZD}  =  \frac{ \alpha \bigg(1-\sum_i \lambda_i  \left(1- P_{FA}\right)^{i}\bigg)}{1-\sum_i \lambda_i \left(1-P_{FA} - (P_{D}-P_{FA}) \sum_j \rho_j (1-\alpha)^{j}\right)^{i}},
\label{Pwzd_irreg}
\end{eqnarray*}

\noindent and \vspace{-.15cm}
\begin{eqnarray*}
P_{ZD} = 1-\sum_i \lambda_i  \bigg( \hspace{-.10cm}1-P_{FA}-  (P_{D}-P_{FA}) \frac{\sum_j j \rho_j (1-\alpha)^{j-1}}{\sum_j j \rho_j}\bigg)^{i}.
\label{pzd_irreg_no}
\end{eqnarray*}

In noiseless scenario, when $P_{D}=1$ and $P_{FA}=0$, from the above equations, we have $P_{WZD}=0$ and $P_{ZD}$ reduces to (\ref{Pzd_irreg_wn}).

\subsection{ Complexity of the Proposed Algorithm}

In terms of the complexity, the proposed scheme has very low complexity as it only needs to identify the zero measurement nodes and verify the adjacent variable nodes as vacant sub-channels. This is negligible in comparison with any of the existing algorithms for the recovery of block sparse signals. In particular, one of the most powerful algorithms for the recovery of block sparse signals, in terms of performance/complexity
trade-off, is AMP with James-Stein's estimator (AMP-JS) \cite{DJM-2013}. In terms of complexity, the complexity of AMP-JS is $\mathcal{O}(MN)$, which is much higher than the minimal complexity of the proposed scheme.

\vspace{-.2cm}
\section{Simulation Results}
\label{sec:Simulations}

In this section, we present some simulation results to demonstrate the effectiveness of the proposed method for detecting the spectrum holes in both noiseless and noisy cases.
We consider an $N$ dimensional signal with the positive frequency range of $[0:N/2]$, containing $L$ non-overlapping channels of equal bandwidth $\frac{N}{2L}$. 
We use an $N$-point unitary Fourier matrix to map the input signal from the time to the frequency domain.
To model the block sparse spectrum, each block is independently selected to be non-zero with probability $\alpha$
(from positive frequency elements). We assign the real and imaginary parts of the elements of non-zero blocks from a Gaussian distribution $\mathcal{N}(0,\sigma_s^2)$ in an i.i.d. fashion. In our simulations, we set $\sigma_s=1$. It is important to note that the choice of these continuous distributions does not affect the performance of the ZMD method, in either noiseless or noisy scenarios.

\vspace{-.2cm}
\subsection{Noiseless Measurements}
\begin{figure}[t]\vspace{-.1cm}
\centerline{\includegraphics[width=3.3in]{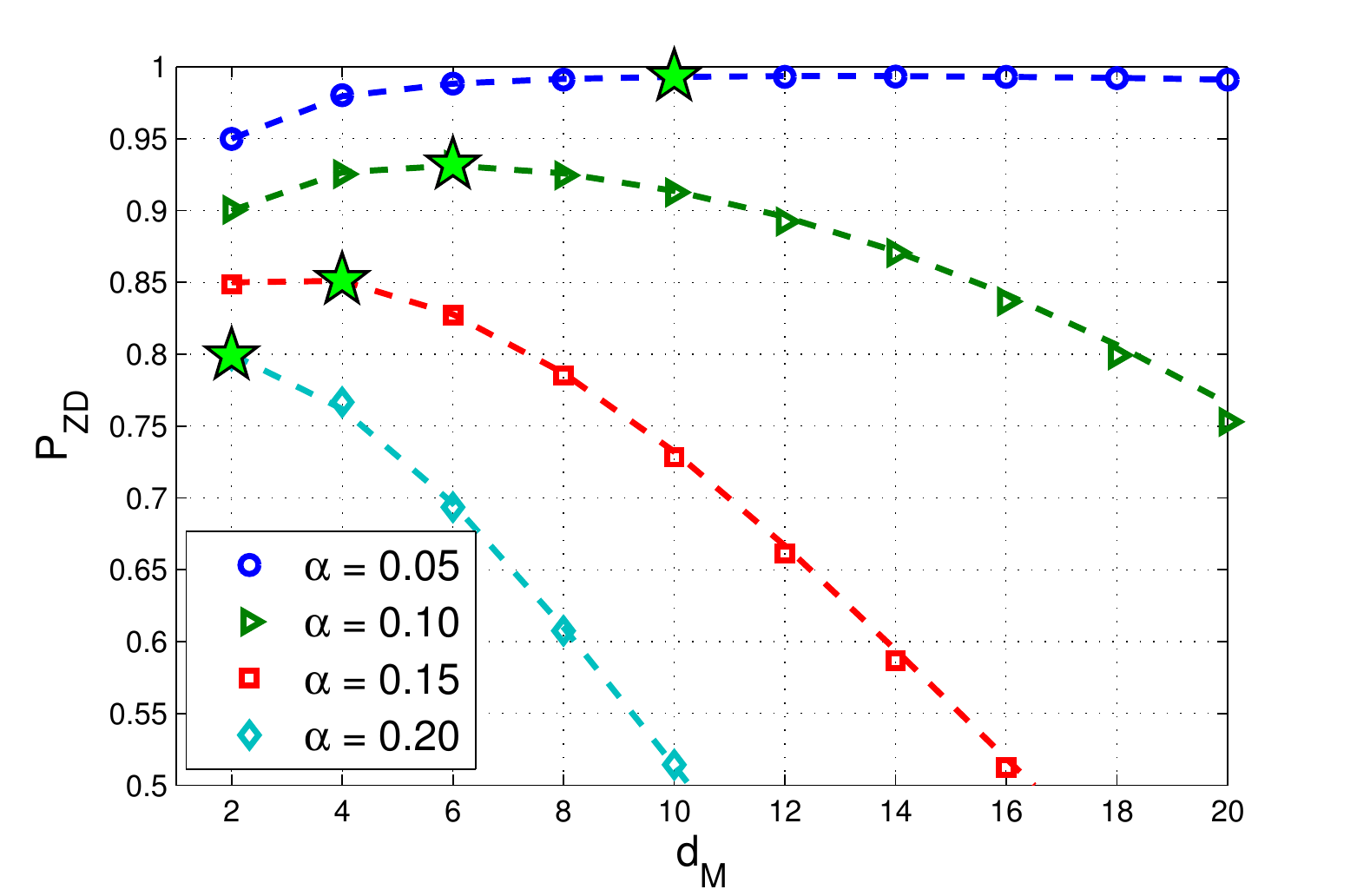}\vspace{-.1cm}}
\caption{{\small $P_{ZD}$ obtained by both analysis (dashed lines) and simulations (solid lines) versus $d_M$ for different sparsity ratios for bi-regular graphs with $L=1000$ and $M=500$. The stars are the maximum $P_{ZD}$ for each sparsity ratio.}\vspace{-.1cm}}
\label{fig:Pzd_vs_dM}
\end{figure}

Fig. \ref{fig:Pzd_vs_dM} shows $P_{ZD}$, obtained both analytically and by simulations, versus $d_M$ for different sparsity ratios. In this figure, $P_{ZD}$ of ZMD is depicted for the regular graphs with $L=1000$ and $M=500$. Thus, $d_V=d_M/2$.
As seen from this figure, simulation and analytical results match perfectly. Furthermore, for each sparsity ratio, the optimum $d_M$ is depicted in terms of maximizing $P_{ZD}$. We see that for smaller sparsity ratios, larger $d_M$ is required to maximize $P_{ZD}$ and with increasing the sparsity ratio, the optimum value of $d_M$ decreases. In particular, for larger sparsity ratios, the best degree distribution is $(d_V,d_M)=(1,2)$.

\begin{figure}[t]\vspace{-.1cm}
\centerline{\includegraphics[width=3.3in]{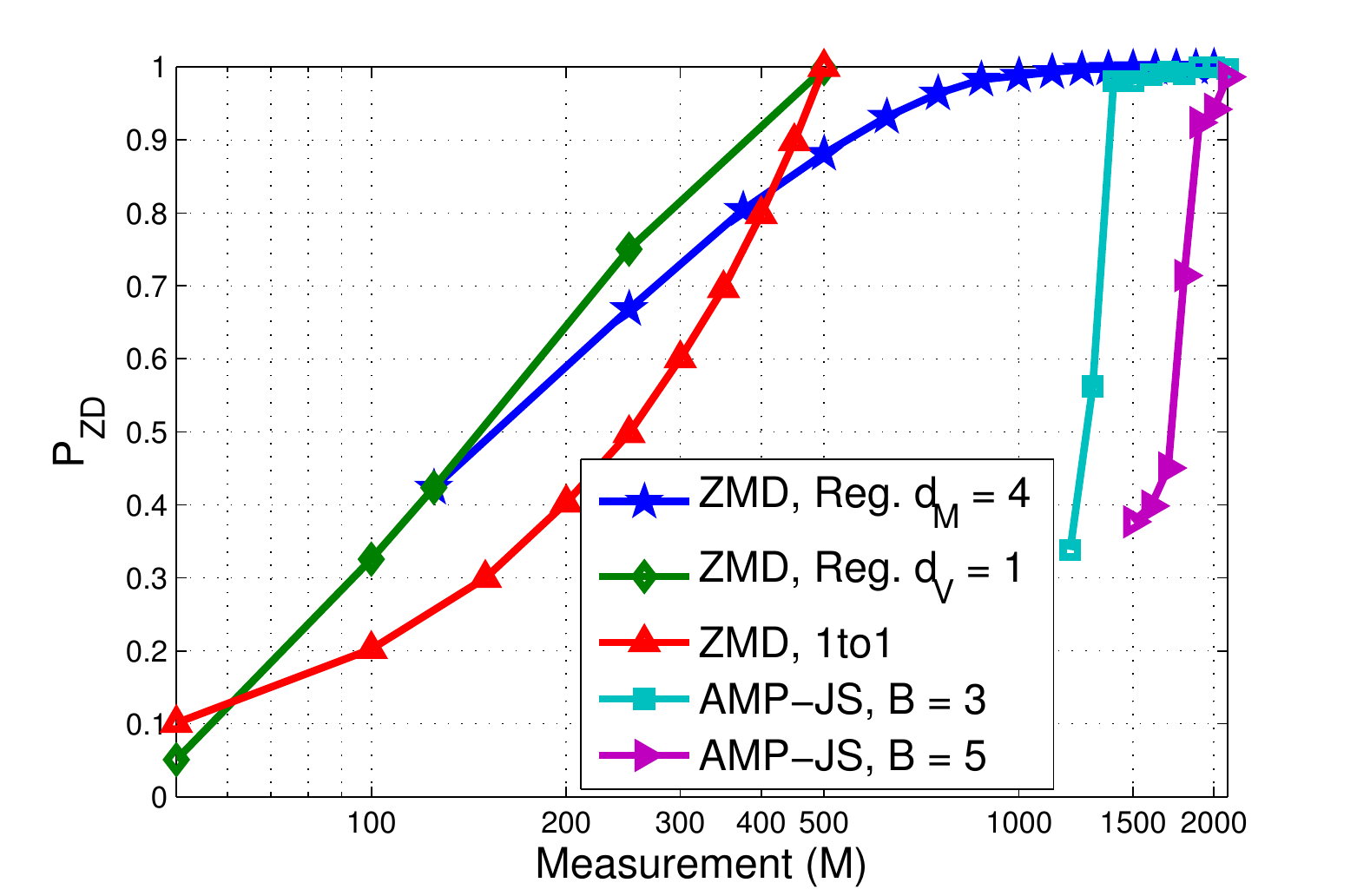}\vspace{-.1cm}}
\caption{{\small $P_{ZD}$ versus $M$ for a signal with $\alpha = 0.25$ and $L=500$, for different algorithms and sensing graphs.}\vspace{-.1cm}}
\label{fig:Pzd_vs_M_algs}
\end{figure}

Fig. \ref{fig:Pzd_vs_M_algs} shows $P_{ZD}$ versus number of measurements, $M$, for different algorithms in a case with $L=500$ and $\alpha=0.25$. Note that the x-axis is in log scale. In this figure, we see $P_{ZD}$ of the regular graphs with $d_V=1$ when $d_M$ is obtained by $d_V L/M$ for each $M$, as well as the regular graphs with $d_M=4$ when $d_V$ is obtained by $d_M M/L$ for each $M$.
We have also plotted the curve corresponding to a graph, where each measurement node is connected to only one variable node and each variable node is connected to at most one measurement node. This case is denoted by ``1to1'' in Fig. \ref{fig:Pzd_vs_M_algs}.

In Fig. \ref{fig:Pzd_vs_M_algs}, we have also plotted $P_{ZD}$ of AMP-JS, which is known as the-state-of-the-art in block sparse signal recovery, for two different block lengths 3 and 5.
For AMP-JS, the sensing matrix is selected to be a dense matrix randomly constructed with i.i.d. elements, where each element is a zero-mean Gaussian random variable with variance $1/M$.
In the AMP-JS, there is a thresholding function based on the $\ell_2$-norm of each block. Thus, the zero and non-zero blocks can be detected easily.
Since in the noiseless scenario, unlike the case for the proposed ZMD scheme, the $P_{WZD}$ of AMP-JS is not necessarily zero, $P_{ZD}$ of AMP-JS is plotted only for the $M$ values in which $P_{WZD}<0.1\%$.
We see that with increasing the block length, the required number of measurements by AMP-JS is increased for a target $P_{ZD}$. However, the performance of the proposed method does not depend on the block length.
The comparison shows the superior performance of the proposed scheme, particularly for regular graphs with $d_V=1$. For very low number of measurements, the 1to1 graph outperforms the others.

\vspace{-.01cm}
\subsection{Noisy Measurements}
\begin{figure}[t]\vspace{-.1cm}
\centerline{\includegraphics[width=3.3in]{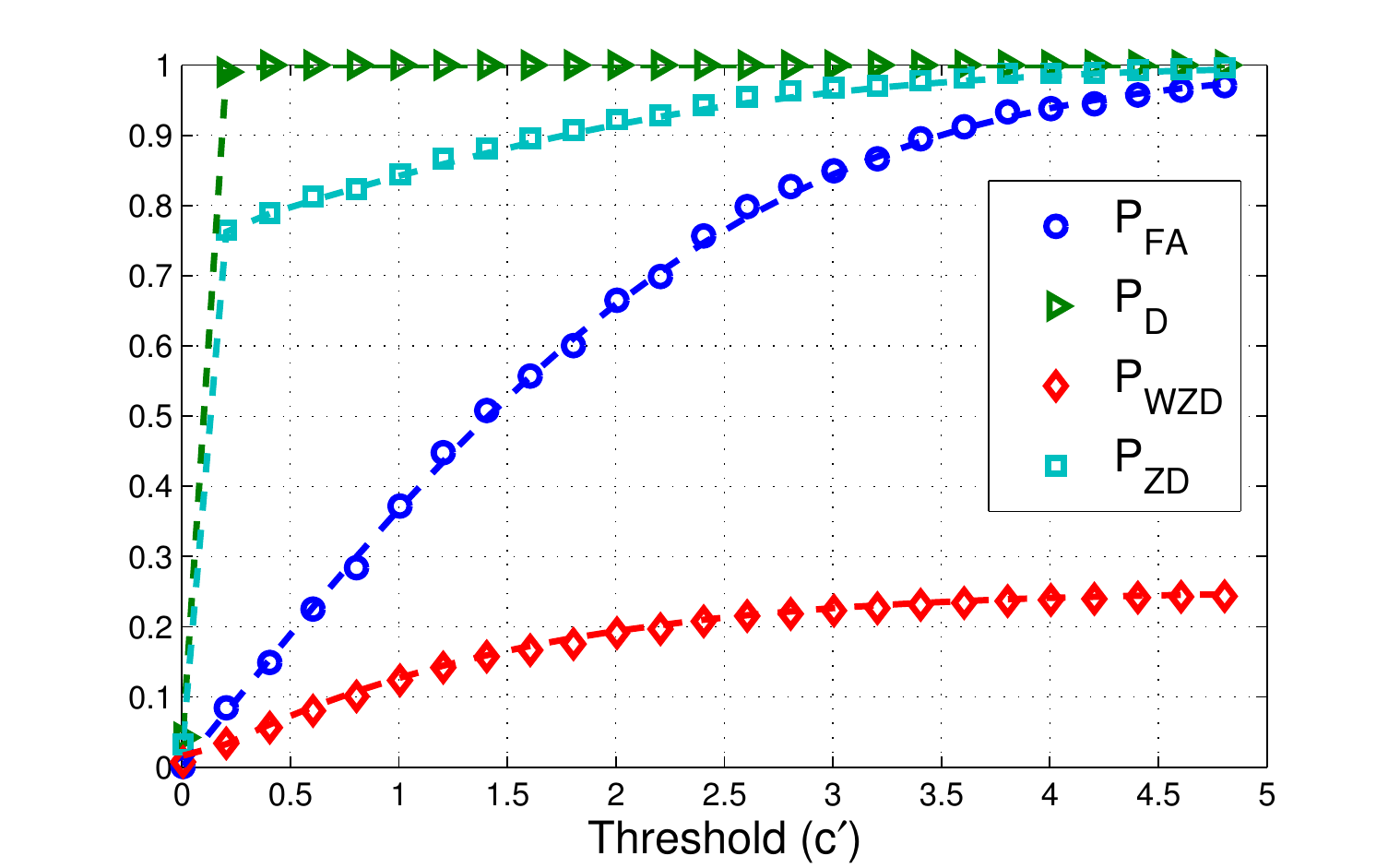}\vspace{-.1cm}}
\caption{\small{The probabilities $P_{D}$, $P_{FA}$, $P_{ZD}$ and $P_{WZD}$ versus the threshold, $c'$, when the sensing graph is regular with degree distribution $(1,2)$ for $L=1000$, $\alpha = 0.25$ and SNR$=25$ dB. The markers and the dashed lines correspond to the simulation and analytical results, respectively.}\vspace{-.1cm}}
\label{fig:Prob_thresh}
\end{figure}

In Fig. \ref{fig:Prob_thresh}, the probabilities of false alarm ($P_{FA}$), detection ($P_{D}$), wrong zero detection ($P_{WZD}$) and zero detection ($P_{ZD}$) are plotted versus the threshold $c'$. Both analytical and simulation results are presented.
The sensing graphs are regular graphs with degrees $d_V=1$ and $d_M=2$. 
The number of blocks $L$ is 1000 and $\alpha = 0.25$, and the curves are plotted for SNR$= 25$ dB. As seen in this figure, the simulation and analytical results match closely.

\begin{figure}[t]\vspace{-.1cm}
\centerline{\includegraphics[width=3.3in]{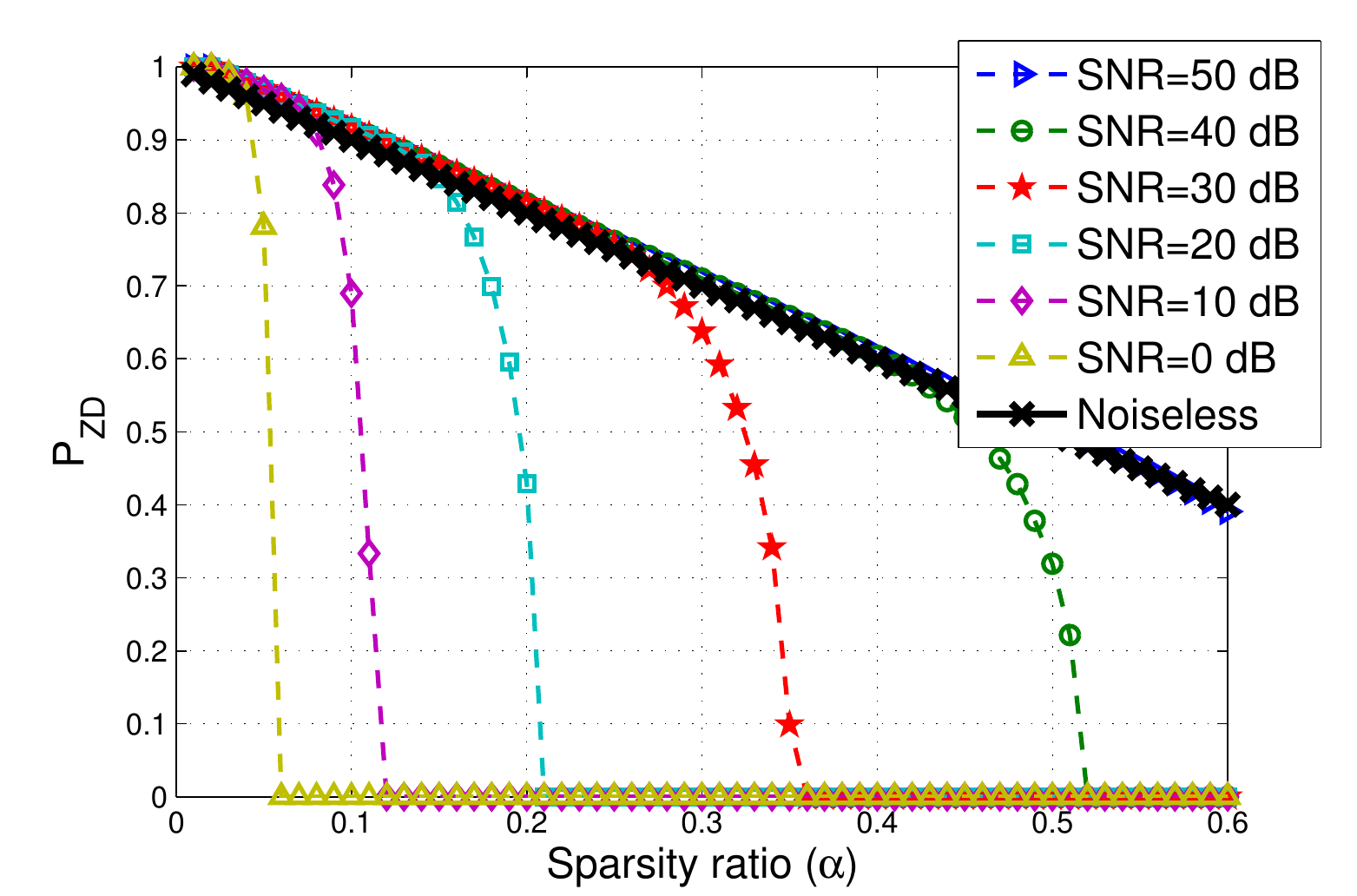}\vspace{-.1cm}}
\caption{\small{$P_{ZD}$ versus sparsity ratio for different noise levels with regular sensing graphs with the degree distribution of (1,2) for $L=1000$ and $M=500$.}\vspace{-.2cm}}
\label{fig:Pzd_vs_sp_4SNRs}
\end{figure}

Fig. \ref{fig:Pzd_vs_sp_4SNRs} shows $P_{ZD}$ versus sparsity ratio for different noise levels when $P_{WZD}$ is limited below 2\%. The curve for the noiseless scenario is also provided for reference. The sensing graphs are regular with degree distribution equal to (1,2), $L=1000$ and $M=500$.
Fig. \ref{fig:Pzd_vs_sp_4SNRs} shows that the performance improves with increasing SNR, and at each SNR, the $P_{ZD}$ drops rapidly when the sparsity ratio in increased beyond a certain threshold. Before such a threshold is reached, the performance is practically the same as that of the noiseless case, but, when this threshold is passed, the $P_{ZD}$ curve demonstrates a waterfall behavior down towards $P_{ZD}=0$. This waterfall region (for a given SNR) corresponds to the values of $\alpha$ for which the threshold $c'$ has to be decreased rapidly (with increasing $\alpha$) to maintain $P_{WZD}$ below 2\%. As a consequence, $P_{ZD}$ is decreased sharply.

\vspace{-.01cm}
\section{Conclusion}
\label{sec:Conclusion}
A novel zero-block detection scheme in the context of wideband spectrum sensing was proposed. For this, an analog to information converter (AIC) with block sparse sensing matrix was designed.
Analytical and simulation results for both noiseless and noisy scenarios were presented. The results demonstrated the effectiveness of the proposed scheme in reliable detection of spectrum holes with minimal complexity even in scenarios where the number of measurements were relatively small. In particular, it was shown that, at the presence of noise, performance similar to the noiseless case can be achieved at higher SNR values and smaller sparsity ratios. Probably one of the most important contributions of this work was to introduce a new compressed sensing paradigm in which one is interested in reliable detection of (some of the) zero blocks rather than the recovery of the whole block sparse signal.
Future work includes the optimization of irregular degree distributions for sensing matrix (graph) to maximize $P_D$ for a given $P_{WZD}$.

\end{document}